# Negative Matter, Repulsion Force, Dark Matter, Phantom and Theoretical Test

## ——Their Relations with Inflation Cosmos and Higgs Mechanism

Yi-Fang Chang
Department of Physics, Yunnan University, Kunming, 650091, China
(e-mail: yifangchang1030@hotmail.com)

**Abstract**: First, dark matter is introduced. Next, the Dirac negative energy state is rediscussed. It is a negative matter with some new characteristics, which are mainly the gravitation each other, but the repulsion with all positive matter. Such the positive and negative matters are two regions of topological separation in general case, and the negative matter is invisible. It is the simplest candidate of dark matter, and can explain some characteristics of the dark matter and dark energy. Recent phantom on dark energy is namely a negative matter. We propose that in quantum fluctuations the positive matter and negative matter are created at the same time, and derive an inflation cosmos, which is created from nothing. The Higgs mechanism is possibly a product of positive and negative matter. Based on a basic axiom and the two foundational principles of the negative matter, we research its predictions and possible theoretical tests, in particular, the season effect. The negative matter should be a necessary development of Dirac theory. Finally, we propose the three basic laws of the negative matter. The existence of four matters on positive, opposite, and negative, negative-opposite particles will form the most perfect symmetrical world.

## 1. Introduction

The speed of an object surrounded a galaxy is measured, which can estimate mass of the galaxy. Many results discover that the total mass of galaxies is always far larger than luminous mass of these galaxies. This shows the existence of dark matter in various galaxies. Dark matter is fundamentally different from the normal matter. It is invisible using modern telescopes because it gives off no light or heat, and appears to interact with other matter only gravitationally. In contrast, the luminous matter is everything commonly associated with the universe: the galaxies, stars, gas and planets. Further, this is confirmed that there are abundant dark matters by the mass-to-light ratio, etc., in group of galaxies and cluster of galaxies, in the universe. And the ratio between dark matter and luminous matter increases with dimensions of these systems [1]. Now investigation of dark matter is a focus of fundamental interest to astronomers, astrophysicists, cosmologists, and nuclear and particle physicists [2].

Since 1981, dark energy is proposed in order to explain the acceleration of inflation in the universe, which is produced due to dark energy as a huge repulsive force [3]. Usually assume that dark energy is a scalar field, and connects with the cosmological constant $\Lambda$, which corresponds to a repulsive force predicted by the general relativity. It is better model that there are about 70% mass in the universe from the cosmological constant. This can explain not only the acceleration of inflation in the universe, and unify many different results of observations.

At present in the universe dark matter has about 24% and dark energy has about 72%, only 4% is visible matter. Dark matter is possibly the weakly interacting massive particle (WIMP), neutrino with mass, baryonic dark matter and nonbaryonic dark matter [4,5], monopole, supersymmetric dark matter [6], axion [7] and so on. Cosmologist divides the candidate of dark matter into three types: hot, warm and cold dark matter. Dark energy seems to be the energy of vacuum.

So far scientists have already done many experiments to try to search the existence of dark matter. Dark matter cannot be observed directly, but it may interfere in light wave or gravitation and so on, emitted from star.

Physicists proposed some new theories, which do not need dark matter, for example, the modified Newtonian dynamics (MOND) [8]. Bekenstein researched the tensor-vector-scalar theories (TeVeS). Mannheim developed the conformal gravity theory [9,10]. It is a geometrized gravitational theory, which possesses gravitation each other for local scale and repulsion each other for cosmological scale. This does not need dark matter and dark energy, and may explain the acceleration of inflation in the universe and data of the circular speed curves of galaxies. Drummond presented a bimetric theory of gravity containing a length scale of galactic size [11]. For distances less than this scale the theory satisfies the standard tests of general relativity, and for distances greater than this scale the theory yields an effective gravitational constant much larger than the locally observed value of Newtonian constant.

The Scientists proposed two different concepts: dark matter and dark energy, whose reason is both different exhibitions. Dark matter seems to have mass and may become huge conglomeration. Cosmologists compute that the gravitational conglomeration of these dark matters is a key function for the process formed galaxies from general matter. But dark energy seems to be zero mass, and distributes uniformly in the whole space, and its interactions are repulsive.

First theoretical model described dark energy is the modified general relativity, which introduces the cosmological constant [12], which can explain many effects of dark energy, but cannot explain dark matter. The cosmological constant may assort the density $\Omega_0 \approx 0.2$ determined by cosmological tests and results predicted by the inflation theory. The inflation theory satisfies a relation $\Omega_0 = 1 - (\Lambda / 3H_0^2)$. When both is equal, so $\Lambda = 2.5 \times 10^{-35} h^2 s^{-2}$ [1]. It is consistent with observational results.

Recently, based on observations of a remarkable cosmic structure called the bullet cluster, Bradac, et al., discovered that this structure is actually two clusters of galaxies passing through one another [13]. As the two clusters cross at a speed of 10 million miles per hour, the luminous matter in each cluster interacts with the luminous matter in the other cluster and slows down. But dark matter in each cluster does not interact, passing right through without disruption. This difference in interaction causes dark matter to sail ahead of the luminous matter, separating each cluster into two components: dark matter in the lead and luminous matter lagging behind. They discovered four separate clumps of matter: two large clumps of dark matter speeding away from the collision, and two smaller clumps of luminous matter trailing in their wake. The spatial separation of the clumps proves that two types of matter exist, while the extreme difference in their behavior shows the exotic nature of dark matter. Past observations have shown that only a very small percentage of mass in the universe can be explained by regular matter. The new research is the first to detect luminous matter and dark matter independent of one another, with the luminous matter clumped together in one region and dark matter clumped together in another. These observations demonstrate that there are two types of matter: one visible and one invisible.

Maggiore and Riotto derived excursion set theory from a path integral formulation, and discussed the classic method for computing the mass function of dark matter halos [14]. Lackner and Ostriker investigated dissipational versus dissipationless galaxy formation and dark matter content of galaxies [15]. Zhao and Li proposed a unified framework for dark matter like modified Newtonian dynamics gravity and dark energy like f(R) gravity, which allows the expanding physical metric to be bent by a single new species of dark fluid flowing in spacetime [16]. Abdo, et al., reported the observations of 14 dwarf spheroidal galaxies with the Fermi Gamma-Ray Space Telescope, and provided a new opportunity to test particle dark matter models [17]. Faltenbacher, et al., analyzed the distribution of dark matter and semianalytical galaxies in the Millennium Simulation to investigate two critical unanswered questions: the workings of nurture effects on galaxies and the whereabouts of the missing baryons, and studied the corresponding halo mass function [18]. Morandi, et al., presented the determination of the intrinsic three-dimensional structure and the physical parameters of both dark matter and intracluster medium in the triaxial galaxy clusters, and solved the long-standing discrepancy between galaxy cluster masses determined from X-ray and gravitational-lensing observations, and supported the cold dark matter scenario [19]. Holley-Bockelmann, et al., calculated the gravitational wave signal from assembling the lightest supermassive black holes, and used high-resolution cosmological N-body simulations to track the merger history of the host dark matter halos [20]. Park, et al., extended the concept of galaxy environment from the local galaxy number density to the gravitational potential and its functions, and inspected the dependence of dark matter halo properties on four environmental parameters. This is evident that the internal physical properties of dark halos are controlled by small-scale physics [21]. Thanjavur, et al., discussed the dark matter distribution and the mass-to-light ratios in galaxy groups from combined analysis of strong lensing and galaxy dynamics [22]. Lundgren, et al., discussed the thermal evolution and Bose-Einstein condensation of ultralight dark matter particles at finite, realistic cosmological temperatures [23]. Hester and Tasitsiomi studied the dependence on environment in the dark matter halo mergers [24].

Scherrer proposed a new k-essence models in which the Lagrangian p is a function only of the derivatives of a scalar field. In the model the universe fills a kind of invisible fluid, and the models can serve as a unified model for dark matter and dark energy [25]. Soleng discussed dark matter and non-Newtonian gravity from general relativity coupled to a fluid of strings [26]. But, the tests of some theories are very difficult.

**2. Rediscussion of negative energy state**

It is well-known that Dirac predicted anti-particles from its equation and the negative energy state [27], and he emphasized: we cannot ignore the negative energy states [28]. In order to prevent to jump continuously from positive energy state to negative energy state in the quantum theories, and keep the stability of world, Dirac proposed that as long as suppose that all the states of negative energy are occupied except perhaps a few of small velocity. The vacuum of the realistic world has already been filling with all negative energy state, such the Pauli exclusion principle will come into play and prevent more than one electron going into any one state, and avoid this jumping difficulty. It is namely the well-known Dirac negative energy sea and whose vacancy or hole, which is an anti-particle (or opposite particle). From this the annihilation and creation between positive and opposite particles may be predicted. There is exact description in <The Principles of Quantum Mechanics> [29]. But, it prevent only jump of fermions, but cannot be applied to bosons. Therefore, the stability problem exists still. The negative energy state appears in all relativity theories, also in the classical theory [28].

The negative energy corresponds to the negative mass, so scientists consider that it will accelerate along a contrary direction under a force according to the Newton second law. Further, Bondi [30] considered three kinds of mass according to the measurement: inertial, passive gravitational, and active gravitational mass, and there are four cases. Here the negative matter responds perversely to nongravitational forces, responds like ordinary matter to gravitational forces, but produces repulsive gravitational fields. But, this is a question: Bondi believes that the positive body will attract the negative one (since all bodies are attracted by it), etc. According to the principle of equivalence in general relativity, inertial mass and gravitational mass should be equal always. Therefore, there are only three cases: positive and positive matters, positive and negative matters, negative and negative matters. In a word, according to the mass-energy relation in Einstein relativity, the Dirac negative energy should correspond to the negative matter.

We think, first, the anti-(opposite) matter and the negative matter should be distinguished exactly. The anti-matter is that some properties of matter are opposite, for instance, charge, baryon number, lepton number, strangeness number and so on, but their masses and total energy are still positive. These particles include positron and various anti-particles. The existence of these particles is already verified. Both positive and opposite matters meet to annihilate to photons or mesons with conservation of energy and zero-charge. The negative matter has a negative mass and total energy. Its main characteristic should be the universal gravitation each other, but is the universal repulsion with all positive matter. Therefore, the creation of negative matter is difficult, but its existence should be also stable. In general case both of positive and negative matters are two regions of topological separation by different interactions. When the positive and negative matters with the same mass meet, they will become a real vacuum. But, so far their existence on the experiment is not final conclusion. Theoretically, in the negative matter there is also negative anti-matter with opposite charge and so on, but with negative mass.

We should extend the Dirac theory, and assume that Dirac sea is in fact a negative matter, and then the anti-particle is only a hole in Dirac sea. Because positive and negative matters are repulsive forces, these holes are stable. Such Dirac sea and its hole theory hold generally for various particles. Generally, a negative energy state must appear in any relativity theory. It is namely the Klein paradox in the relativistic quantum theory.

The Dirac equations of fermions can describe anti-matter. The cosmological constant $\Lambda$ describes possibly the negative matter, which corresponds to the $\Lambda$ term in the gravitational field equation. In the Klein-Gorden equation the $m^2$ term may correspond to $\pm m$, both describe bosons. In the Dirac equations m→-m may also describe the negative matter. A universal relation is:

$$E^2 = m^2 c^4 + c^2 p^2. \tag{1}$$

It may be generally applied for various positive, opposite and negative matters, and for all $\pm m$, $\pm E$ and $\pm p$. Only the mass is negative in the equations described negative matter, while the charge and so on are opposite in the equations described opposite matter. For a relation:

$$E^2 = m^2c^4 + c^2(p - \frac{q}{c}A)^2, \text{ i.e., } p - \frac{q}{c}A = \pm\frac{1}{c}\sqrt{E^2 - m^2c^4}, \qquad (2)$$

$$\therefore q = (mv \mp \frac{1}{c}\sqrt{E^2 - m^2c^4})\frac{c}{A}. \qquad (3)$$

Such the charge may be positive or negative, and is particular distinct for v=0. It corresponds to the opposite matter.

The negative matter is possibly influence on the universal gravitational laws, classical mechanics, the motion laws of planet, electrodynamics, general relativity and quantum mechanics, etc. In this case the light ray is red shift at the neighborhood of a gravitational field (positive matter), and is violet shift at the neighborhood of a repulsive field (negative matter),

$$\Delta\lambda/\lambda = -MG/rc^2. \qquad (4)$$

Of course, light emitted from the negative matter cannot be observed directly. The light ray should have repulsive deflection in a field of the negative matter,

$$\alpha = -4MG/c^2R, \qquad (5)$$

and a more general deflection should be

$$\alpha = 4G(M_1 - M_2)/c^2R, \qquad (6)$$

in which $M_1$ and $M_2$ are mass of the positive and negative matters, respectively.

For the Kepler laws of planet,

$$F(r) = -\frac{G}{r^2}(M_1 - M_2), u = \frac{1}{r}. \qquad (7)$$

So $\qquad \frac{d^2u}{d\vartheta^2} + u = \frac{G}{H}(M_1 - M_2). \qquad (8)$

Its solution is:

$$r = \frac{H/G(M_1 - M_2)}{1 + CH\cos(\vartheta - \vartheta_0)/G(M_1 - M_2)}. \qquad (9)$$

When $\vartheta_0 = 0$, it becomes a quadric curve,

$$r = \frac{p}{1 + e\cos\vartheta}, \qquad (10)$$

in which $e = Cp = CH/G(M_1 - M_2)$. It is ellipse for E<1 and $M_1 > M_2$; it is hyperbola for E>1 and $M_1 < M_2$; it is parabola for E=1 and $M_1 = M_2$. This is a modified Kepler first law. The Kepler second law should be invariable.

In the gravitational law:

$$F = -\frac{G}{r^2}m_1m_2, \qquad (11)$$

there are two masses, but in the Newton second law F=ma there is only one mass. In order to keep the consistency of natural laws, and a repulsive force between positive and negative matters, we should suppose -F=-m*a*, i.e., F=m*a* hold always for the negative matter, so that *a* is still an acceleration in the negative matter, while is always deceleration between the positive and negative matters.

In the special relativity the mass increases still. In the four-vector change only is ($\pm mv$; $\pm E/c$), the time-like interval is -v<-c, i.e., v>c; the space-like interval is -v>-c, i.e., v<c, both are just opposite. Therefore, the superluminal is in the time-like interval. In the general relativity there is similarly curved time-space.

In the quantum mechanics the negative matter may be still E=hv, in which E→-E and -h→h. Such the de Broglie wave length is positive. The uncertainty principle

$$(\Delta x)^2 (\Delta p_x)^2 \geq \hbar^2 / 4 , \tag{12}$$

is invariant, but another relation

$$(\Delta x)(-\Delta p_x) \geq -\hbar/2 , \tag{13}$$

will become probably to

$$(\Delta x)(\Delta p_x) \leq \hbar/2 . \tag{14}$$

The Heisenberg equation is also invariant, mass becomes an opposite sign in the Schrodinger equation, because the energy-momentum operators are invariant. Such

$$-E = \frac{p^2}{(-2m)} + U(r) , \tag{15}$$

whose corresponding equation in quantum mechanics is:

$$i\hbar \frac{\partial \psi}{\partial t} = \frac{\hbar^2}{2m} \nabla^2 \psi - U(r)\psi . \tag{16}$$

Here only U→-U. The Klein-Gordon equation and the Dirac equations are invariant. But, an equation in an electromagnetic field is different:

$$[-E + e\phi - \alpha(-cp + eA) + \beta\mu c^2]\psi = 0 . \tag{17}$$

**3. Negative matter is possibly a dark matter**

A unique dependable method determined all mass is to study their gravitation effect, for which the easiest method is the measurement of the circular speed curves in the galaxy [31]. These curves may be measured from the Doppler shift of spectrum [32].

Dark matter self does not emit light, and does also not interact with light. In the negative matter there is the negative photon, which possesses negative energy and negative mass, and is repulsion with general matter, so the negative matter is invisible. The state equation of dark energy is different with the equation of usual matter, and at present assume that it is repulsive force each other. So this may correspond to the negative matter [33,34]. Moreover, according to the mass-energy relation in Einstein relativity, dark matter and dark energy should be unified in this case.

Recently, Caldwell proposed phantom as cosmological consequences of a dark energy component with super-negative equation of state, whose cosmic energy density has negative pressure [35]. Then phantom becomes an important dark energy model [36-60], where the kinetic energy is negative. Such it must possess negative mass according to classical mechanics or relativity, and is namely the negative matter. Hong, et al., considered a higher dimensional cosmological model with a negative kinetic energy scalar phantom field and a cosmological constant [40]. Scherrer and Sen examined phantom dark energy models produced by a field with a negative kinetic term [44]. Chimento, et al., discussed the dark energy density derived from the 3-scalar phantom field, and its negative component plays the role of the negative part of a classical Dirac field [48]. Gonzalez and Guzman presented the first full nonlinear study of a phantom scalar field accreted into a black hole. Here the analysis includes that the total energy of the space-time is positive or negative [53].

The observations for luminous mass find that the velocity V is approximately constant, for example, in a range of radio 0.5 kpcs<R<20 kpcs for our Galaxy [61]. This is an important proof of the existence of dark matter, and which exists in the galactic halo. For a galaxy, if the movement of a star round the center of the galaxy obeys the Kepler law, and the negative matter is introduced, the equation of the star with mass m and distance R to the center will be

$$\frac{Gm}{R^2}(M_1 - M_2) = \frac{m}{R}V^2. \qquad (18)$$

The total mass of this galaxy inside radius R is:

$$M(R) = M_1 + M_2 = \int_0^R \rho(r)dV = \int_0^R \rho(r)4\pi r^2 dr. \qquad (19)$$

Such $\frac{dM(r)}{dr} = 4\pi r^2 \rho(r)$. The continuity equation is:

$$\frac{\partial \rho}{\partial t} + \nabla(\rho v) = 0, \qquad (20)$$

in which $\rho = \rho_1 + \rho_2$ is a total density. The Euler equation is:

$$(\rho_1 + \rho_2)\frac{dv}{dt} = (\rho_1 + \rho_2)[\frac{\partial v}{\partial t} + (v \cdot \nabla)v] = -\nabla(p_1 + p_2) - (\rho_1 + \rho_2)\nabla\Phi. \qquad (21)$$

The cosmological constant corresponds to a fictitious fluid introduced, whose density is $\rho_\Lambda = \Lambda/8\pi G$, and pressure is $p_\Lambda = -\Lambda c^2/8\pi G$. The mass-to-light ratio connects to $(\rho_1 + \rho_2)/\rho_1 = 1 + \rho_2/\rho_1$, such more is the negative matter, and bigger is the mass-to-light ratio.

From Eq.(18), we may obtain a radial velocity

$$V = \sqrt{\frac{G(M_1 - M_2)}{R}}, \qquad (22)$$

of the star, and an angle velocity

$$\Omega = \sqrt{\frac{G(M_1 - M_2)}{R^3}}, \qquad (23)$$

from the movement equation. Such this measurement determines only difference of positive mass and negative mass, i.e., a breaking part of symmetry between positive and negative matters.

We suppose an isolated particle system with the positive and negative matters under gravitational self interaction, whose kinetic energy:

$$T = T_1 - T_2 = \frac{1}{2}(\sum_i m_i \dot{r}_i^2 - \sum_j m_j \dot{r}_j^2). \tag{24}$$

It is simplified to

$$T = \frac{1}{2}(M_1 - M_2)<V^2> = \frac{3}{2}(M_1 - M_2)<V_{saw}^2>. \tag{25}$$

For an object of spherical symmetry the potential energy is:

$$U = -\frac{G}{R}(M_1 - M_2)^2 = -\frac{G}{R}(M_1^2 + M_2^2 - 2M_1 M_2). \tag{26}$$

Applied the virial theorem determined the mass of cluster of galaxies, the sum of kinetic energy T and potential energy U for this system is:

$$2T + U = \frac{1}{2}\frac{d^2}{dt^2}(\sum_i m_i r_i^2 - \sum_j m_j r_j^2) = 0. \tag{27}$$

The kinetic energy of entire system in each particle as a galaxy is namely T. By above formula, the mass of this galaxy becomes:

$$M_1 - M_2 = \frac{3R}{G}<V_{saw}^2>. \tag{28}$$

Therefore, the existence of the negative matter will derive bigger decrease of mass by this way. For example, assume that the positive matter and negative matter are 55% and 45% of total mass, respectively. We observed mass that is only its 10%. Such the negative matter is possibly an important reason produced an effect of dark matter. Moreover, the negative matter is repulsive force for photon, and negative-photon is also repulsive force for matter, both cannot be observed, and show dark matter.

The field equations of general relativity on the negative matter are:

$$G_{\mu\nu} = 8\pi k(T_{\mu\nu} - T'_{\mu\nu}), \tag{29}$$

i.e., $$G_{\mu\nu} + 8\pi k T'_{\mu\nu} = 8\pi k T_{\mu\nu} = G_{\mu\nu} + \Lambda g_{\mu\nu}. \tag{30}$$

So $\Lambda$ corresponds to the negative matter. And

$$\Lambda = 8\pi k T'_{\mu\nu}/g_{\mu\nu} = [\rho' + (p'/c^2)](u_\mu u_\nu / g_{\mu\nu}) - p'. \tag{31}$$

On the other hand, the gravitational field equation with the cosmological constant is extended to:

$$G_{\mu\nu} = 8\pi k T_{\mu\nu} \Rightarrow 8\pi k(T_{\mu\nu} + \Lambda g_{\mu\nu}). \tag{32}$$

Here $\Lambda g_{\mu\nu}$ corresponds to the negative energy state and vacuum energy, i.e., Dirac sea. The Friedmann equation is:

$$\ddot{R}(t) = -\frac{4}{3}\pi G[(\rho_1 - \rho_2) + 3(p_1 - p_2)/c^2]R(t), \tag{33}$$

in which $(\rho_1 - \rho_2) + 3(p_1 - p_2)/c^2$ is effective mass density.

$$\dot{R}^2 - \frac{8}{3}\pi G(\rho_1 - \rho_2)R^2 = 2C, \tag{34}$$

in which $\dot{R}(t_0) = H_0$ is the Hubble constant. The density parameter is:

$$\Omega_0 = \frac{8\pi G \rho_0}{3H_0^2} = \frac{\rho_0}{\rho_c}, \tag{35}$$

$\rho_0 = \rho_1 - \rho_2$ is an observed density. The accelerating expansion of the universe shows

$$(\rho_1 - \rho_2) + 3(p_1 - p_2)/c^2 < 0, \quad \text{i.e.,} \quad \rho_2 + (3p_2/c^2) > \rho_1 + (3p_1/c^2). \tag{36}$$

The negative matter is more than the positive matter.

For the negative matter there should also have the corresponding black hole, whose radius is:

$$r = -2Gm/c^2. \tag{37}$$

Various positive matter and black hole exhibit the gravitational lensing effect. While the negative matter and negative black hole will be the repulsive lensing phenomena. Both should be different in observations.

**4. Negative matter and inflationary cosmology**

In the standard model of hot big-bang cosmology there are some problems [62,63], for example, the horizon problem, the flatness problem, the antimatter problem, the structure problem and the expansion problem, etc. Such Guth proposed an inflationary universe. In the early history the universe pass through some phase transitions when the latent heat is released. A huge expansion factor results from a period of exponential growth. It is based on the grand unified models (GUM) of particle interactions [62], and the Higgs field [63]. The inflationary universe may explain some problems and the monopole problem. Further, the inflation theory is extended to the chaotic inflation models [64,65]. These inflation theories relate dark matter [3].

Since a Universe must have a zero net value for all conserved quantities, and it must consist equally of matter and anti-matter, Tryon supposed that the Universe is a quantum fluctuations in vacuum [66], and the creation of the cosmos from nothing [66,67].

We propose that under this case the positive matter and negative matter are created at the same time. It is a Planck time, whose time scale is about $10^{-43}$ s, and whose length is about $10^{-33}$ cm. At this very small space the positive matter and negative matter are repulsive each other, and are the very strong repulsive interaction, whose ratio with the gravitational interaction is $15/10^{-39} \sim 10^{40}$. Therefore, the Universe inflates, which is a phase transformation of the Grand Unified Theories (GUT) at the form of the strong interaction. While an exponential inflation is just a form of the strong interaction:

$$F = -g^2 \frac{e^{-kr}}{r^2}. \qquad (38)$$

Here the positive matter is g, and the negative matter is –g, so F>0 is a huge strong repulsive force for the length inside $10^{-13}$ cm. When the time is $10^{-34}$ s and the length is bigger than one of the strong interaction, the inflation finishes, and the positive matter and opposite matter are created. While the force between the positive matter and negative matter will become a usual repulsion. When the space between the positive matter and negative matter are bigger, both will form two regions of topological separation repulsed each other.

A false vacuum becomes two true vacuums in the inflation [63], namely, a pair of worlds on the positive matter and negative matters, in which the negative matter is dark matter. This corresponds to the cosmological mode created from nothing to all things. It may form the parallel worlds, or the many-worlds, or multiverse, etc.

L.Ford (1978) proposed that the negative energy must obey the three theorems. But they are completely different with the negative matter proposed in this paper. In a discussion on the time-enginery K.Thorne, M.Morris and U.Yurtsever (1988) proposed also the negative matter and negative energy, which and the positive matter and negative matter repulse each other. But, usual consideration is that the negative matter cannot exist. While only the negative energy exists, for example, the energy of the gravitation is negative, and in the Casimir effect and so on. S.Hawking proposed also that in order to stabilize all solutions of wormhole, the negative matter is necessary. The negative matter is also called an exotic matter.

The Schwarzschild metric of the negative matter should be [68]

$$ds^2 = (1 + \frac{2m}{r})dt^2 - \frac{dr^2}{1+(2m/r)} - r^2(d\theta^2 + \sin^2\theta d\varphi^2). \qquad (39)$$

When Einstein and Rosen investigated the particle problem in the general relativity, a new variable $u^2 = r + 2m$, i.e., $u = \pm\sqrt{r+2m}$ is introduced. Then the two congruent parts or sheets corresponding to u>0 and u<0, which are joined by a hyperplane (Einstein-Rosen bridge) [68] $r = -2m$ or u=0, in which $g_{\mu\nu}$ vanishes.

Pachner studied the nature of singularities in relativistic world models [69]. It is shown that a discontinuous inversion of velocities of particles at the moment of maximum contraction of Friedman universes is removed by the assumption that the square root of the reciprocal absolute value of the Riemannian curvature of the three-dimensional space takes on not only positive, but also negative values. Albrecht, et al., estimated the power spectra of density fluctuations produced by cosmic strings with neutrino hot dark matter (HDM), and found that the spectrum has more power on small scales than inflation, less than cold dark matter (CDM) inflation [65].

Recently, Henriques, et al., investigated a cosmological model, based on the Salam-Sezgin six-dimensional supergravity theory. Assuming a period of warm inflation, they shown that it is possible to extend the evolution of the model back in time, to include the inflationary period, thus unifying inflation, dark matter, and dark energy within a single framework. It shows that present-day theories, based on superstrings and supergravity, may eventually lead to a comprehensive modeling of the evolution of the universe [70]. Hohmann and Wohlfarth argued that the most conservative geometric extension of Einstein gravity describing both positive and negative mass sources and observers is bimetric gravity and contains two copies of standard model matter which interact only gravitationally. Matter fields related to one of the metrics then appear dark from the point of view of an observer defined by the other metric, and so may provide a potential explanation for the dark universe. In this framework they considered the most general form of linearized field equations compatible with physically and mathematically well-motivated assumptions. Using gauge-invariant linear perturbation theory, they proved a no-go theorem ruling out all bimetric gravity theories that, in the Newtonian limit, lead to precisely opposite forces on positive and negative test masses [71]. This proves just the basic repulsion of my negative matter.

**5.Negative matter and Higgs mechanism**

The Higgs field is necessary, from this the symmetries are spontaneously broken, and the gauge bosons obtain masses [72-74]. The Higgs field equations are [74]:

$$\nabla_\mu \nabla^\mu \varphi_a + \frac{1}{2}(m_0^2 - f^2 \varphi_b \varphi_b)\varphi_a = 0. \tag{40}$$

If $\varphi_b = 0$ or $f=0$, so $m_0^2 < 0$ for usual field. Such Higgs boson cannot be measured, and so far various experiments do not search any Higgs boson. It is a puzzle that the Higgs bosons form various mesons and baryons whose masses are very small than huge Higgs mass along with the energy increased by those accelerators.

The Lagrangian of two scalar or pseudo-scalar fields without spin is [75]:

$$L = \frac{1}{2}(\partial_\mu \varphi_1)^2 + \frac{1}{2}(\partial_\mu \varphi_2)^2 - \frac{1}{2}\mu^2(\varphi_1^2 + \varphi_2^2) - \frac{\lambda}{4}(\varphi_1^2 + \varphi_2^2)^2. \tag{41}$$

It is an ordinary solution for $\mu^2 > 0$. While $\mu^2 < 0$ corresponds to the Goldstone Lagrangian, which has two solutions:

$$|\varphi| = (\varphi_1^2 + \varphi_2^2)^{1/2} = \begin{cases} 0 (V_s = 0). \\ V_0 = \sqrt{-\mu^2/\lambda} (V_s = -\mu^4/4\lambda). \end{cases} \tag{42}$$

The non-ordinary solution obtains a minimal value $V_s = -\mu^4/4\lambda$ and a non-zero vacuum expectation value, and the symmetries are spontaneously broken. From this various particles obtain masses. It is namely the Higgs mechanism [72-76].

The Lagrangian is [76]:

$$L_{\varphi,QUAD} = -\frac{1}{2}\sum_n [\partial_\mu \varphi'_n - i\sum_{m,\alpha}(t^\alpha_{nm} A_{\alpha\mu} v_m)]^2 = -\frac{1}{2}\sum_n (\partial_\mu \varphi'_n \partial^\mu \varphi'_n) - \frac{1}{2}\sum_{\alpha\beta}(\mu^2_{\alpha\beta} A_{\alpha\mu} A^\mu_\beta). \tag{43}$$

Here the mass matrix

$$\mu^2_{\alpha\beta} \equiv -\sum_{nml} t^\alpha_{nm} t^\beta_{nl} v_m v_l. \tag{44}$$

It should be $\mu^2_{\alpha\beta} < 0$. Further, this yields a ghost Lagrangian.

In Higgs field the $m_0^2 < 0$ originate possibly from a product of positive and negative matter

$(m)(-m) = -m^2$, but it is not an imaginary particle. This show that when the Higgs field is decomposed, the positive matter is namely particle, and obtain spontaneously mass; the negative matter is namely dark matter. The corresponding Goldstone boson is (+m)+(-m)=0, which is a symmetry. While the Higgs mechanism is spontaneously broken symmetry. If the Higgs bosons are tested in search, it will imply that the positive and negative matters are obtained. Generally, the superluminal and corresponding imaginary mass may be also explained by the same way, i.e., a product of positive mass and negative mass.

The Dirac equations of positive matter are:

$$(\gamma_\mu \partial_\mu + m)\psi = 0. \tag{45}$$

The Dirac equations of negative matter are:

$$(\gamma_\mu \partial_\mu - m)\psi = 0. \tag{46}$$

While $\quad (\gamma_\mu \partial_\mu + m)(\gamma_\mu \partial_\mu - m)\psi = (\Box - m^2)\psi = 0. \tag{47}$

It is the Klein-Gordon equation. For

$$(\gamma_\mu \partial_\mu + m)(\gamma_\mu \partial_\mu + m)\psi = (\Box + 2m\gamma_\mu \partial_\mu + m^2)\psi = 0, \tag{48}$$

it is replaced by the Dirac equation, then this is still the Klein-Gordon equation. For the Dirac equations of the negative matter the Klein-Gordon equation is obtained yet by the same method. This shows that there is the same Klein-Gordon equation for the positive-positive matter, or positive-negative matter, or negative-negative matter.

For the Higgs equation and the Klein-Gordon equation only the mass has the opposite sign. Such it may be structured from a pair of the Dirac equations of the positive-negative matter. And it is simplified to the Schrodinger equation, only whose mass is an opposite sign.

In a certain extent an adjoint field of the Dirac field is namely the negative matter, whose equations are:

$$\gamma_\mu \overline{\psi} \partial_\mu - m\overline{\psi} = 0. \tag{49}$$

Here $\overline{\psi} = \psi^+ \gamma_4$, and $\psi^+$ is an adjoint operation, and is a covariant operation [77]. This seems to imply that the negative matter accompanies the positive matter. $\psi\overline{\psi}$ may construct the six types and 16 quantities, in which an antisymmetric tensor is [77]:

$$T_{\mu\nu} = \overline{\psi} \frac{\gamma_\mu \gamma_\nu - \gamma_\nu \gamma_\mu}{2i} \psi. \tag{50}$$

It corresponds perhaps to a repulsive force, and is not a symmetric tensor with 10 components of the gravitational field.

For two components the matrix Dirac equations are [77]:

$$c\sigma_i p_i \psi_S + m_0 c^2 \psi_L = E\psi_L, \tag{51}$$

$$c\sigma_i p_i \psi_L - m_0 c^2 \psi_S = E\psi_S. \tag{52}$$

For (52) there is

$$\psi_S = \frac{c\sigma_i p_i}{E + m_0 c^2} \psi_L, \tag{53}$$

there are the two types and 8 quantities [77]. At low energy $\psi_S$ is a negative mass and a negative energy. When the negative energy cannot be neglected, we derive the Schrodinger equation and the Pauli equation. From (51) there is

$$\psi_L = \frac{c\sigma_i p_i}{E - m_0 c^2} \psi_S. \tag{54}$$

It is replace by Eq.(52), and obtain:

$$c^2(\sigma_i p_i)^2 \psi_S + m_0^2 c^4 \psi_S = E^2 \psi_S. \tag{55}$$

Let $E = -m_0 c^2 + E'$, for non-relativity (low energy) $E' \ll m_0 c^2$, so $E^2 = m_0^2 c^4 - 2m_0 c^2 E'$. Assume

$$(\sigma_i p_i)^2 = -\hbar^2 \Delta, \tag{56}$$

so $\quad -\hbar^2 \Delta \psi_S = -2m_0 E' \psi_S. \tag{57}$

It is the Schrodinger equation of the negative matter. If

$$(\sigma_i p_i)^2 = -\hbar^2 \Delta - (e\hbar/c)\sigma_l H_l, \tag{58}$$

we will obtain the Pauli equation of the negative matter, whose magnetic pole will be $\mu = -e\hbar/2m_0 c$. This is the same results in the Schrodinger equation and the Pauli equation replaced directly by the negative mass.

**6. Theoretical test and the basic laws of negative matter**

The existence of dark matter should affect some results of the Newtonian gravitation and the general relativity, for example, the cosmic average density will increase about 20 times. Dark matter is very complex, and it should include various types, the negative matter is possibly only one of dark matter.

In the large-scale space, if there has a negative matter cluster in the positive matter, a part of positive matter will be screened, and another visible matter changes shape by the repulsive lens. Therefore, the visible matter looks much less. The negative matter and their screening positive matter will exhibit the invisible dark matter. According to this hypothesis, since the screening part and distorted part are different, the star-shape observed will be a little different from different positions of the Earth at the solar system. This season effect may be tested. Mortonson, et al., also proposed testable dark energy predictions from current data [78].

Further, the negative matter can predict: 1. There is repulsive force between positive and negative matters, and which obey the square inverse ratio law according to the Newtonian law. 2. General photons are reflected by negative matter, which exhibits a type of dark matter. 3. Dark matter includes the negative matter and the positive matter screened. 4. When the move speed between positive and negative matter is very big, and the kinetic energy is bigger than the potential energy, their colliding result will be a complete annihilation, whose leftover is only a mass-difference of positive and negative matters. 5. Usual light under interaction of negative matter is repulsive deflection, so it shows the repulsive lensing effect. 6. The negative matter is similar with invisible black hole, but is repulsive force for matter, and its mass is invariant. 7. The negative matter may represent the cosmic repulsion and the fast expansion. 8. The positive and negative matters under some exceeding conditions may be created from nothing at the same time. They will also be main tests of the existence of negative matter.

The negative matter as dark matter will be a kind of the simplest candidate, and can explain some characteristics of a huge lack of mass on dark matter, and of a repulsive force of dark energy. The negative matter determinates the cosmological constant, and changes possibly the gravitational lensing effect, and is consistent with the conformal gravity theory [9,10], and with the observation of the bullet galaxy cluster [13]. The latter shows obviously a huge dissimilarity between the positive and negative matters. In this case two galaxies collide sharply and meet, but the negative and positive matters are repulsive each other, so the negative matter passes very quickly.

The above discussions are based on a basic axiom: the no-contradiction of natural laws, and on the two foundational principles: 1). The negative matter obeys the same natural laws of usual matter, including classical, relativistic and quantum physics. 2). There is universal repulsive force between the positive and negative matter. Of course, some concrete laws are possibly different, for example, the Kepler first law and so on.

Usually, one considers that Dirac predicted the opposite matter. In fact, the Dirac theory implied already the negative matter, which should be also a necessary development of this theory. It is also a development of the Dirac negative energy state. Dirac pointed out: The physical laws are symmetrical between the positive and negative charge. Further, the physical laws should be also symmetrical between the positive and negative matter. It will form just the most perfect symmetrical world that four matters on positive, opposite, and negative, negative-opposite particles exist together.

Finally, we propose particularly the three basic laws or principles of the negative matter: I. The classical law. The negative matter is repulsive with the positive matter, and obeys the Lorentz transformation, etc. II. The quantum law. For the fermions of the positive and negative matter, and corresponding Dirac equations and so on, both masses are opposite; while for the bosons of the positive and negative matter, and corresponding the Klein-Gordon equation and so on, both $m^2$ are the same. III. The symmetry (completeness) law. The physical laws are the most perfect symmetries for four matters of positive, opposite, and negative, negative-opposite particles.

If the negative matter is verified, a new and complete world will be exhibited.